\documentclass[a4paper,oneside,twocolumn]{article}

\usepackage{geometry}
\usepackage{graphicx,graphics,amsmath,amssymb,amsfonts,verbatim, bm, color, mathrsfs, enumerate, amsthm, dsfont, cuted, multirow,authblk,lmodern,abstract,subfig}

\usepackage[normalem]{ulem}
\usepackage[breaklinks=true,colorlinks,citecolor=blue,linkcolor=blue,urlcolor=blue]{hyperref}

\DeclareMathOperator*{\argmin}{arg\,min}

\usepackage[textwidth=2.0cm, textsize=tiny]{todonotes} 

\usepackage[sorting=none,giveninits,style=numeric-comp,doi=false,isbn=false,url=false,date=year,eprint=false,maxnames=5]{biblatex}
\renewbibmacro{in:}{}
\AtEveryBibitem{\clearfield{pages}}
\AtEveryBibitem{\clearfield{number}}

\newcommand*\samethanks[1][\value{footnote}]{\footnotemark[#1]}

\begin{document}

\title{Characterizing  metastable states with the help of machine learning}

\author[1]{Pietro Novelli\thanks{P.N. and L.B. contributed equally to this work.}\footnote{pietro.novelli@iit.it}}
\author[2]{Luigi Bonati\samethanks[1]\footnote{luigi.bonati@iit.it}}
\author[1]{Massimiliano Pontil}
\author[2]{Michele Parrinello\footnote{michele.parrinello@iit.it}}
\affil[1]{Computational Statistics and Machine Learning, Italian Institute of Technology,\newline Via Enrico Melen 83, 16142 Genoa, Italy}
\affil[2]{Atomistic Simulations, Italian Institute of Technology, Via Enrico Melen 83, 16142 Genoa, Italy}
\date{\vspace{-5ex}}

\twocolumn[
  \begin{@twocolumnfalse}
    \maketitle
    \begin{abstract}
    \normalsize Present-day atomistic simulations generate long trajectories of ever more complex systems. Analyzing these data, discovering metastable states, and uncovering their nature is becoming increasingly challenging. In this paper, we first use the variational approach to conformation dynamics to discover the slowest dynamical modes of the simulations. This allows the different metastable states of the system to be located and organized hierarchically. The physical descriptors that characterize metastable states are discovered by means of a machine learning method. We show in the cases of two proteins, Chignolin and Bovine Pancreatic Trypsin Inhibitor, how such analysis can be effortlessly performed in a matter of seconds. Another strength of our approach is that it can be applied to the analysis of both unbiased and biased simulations.
    \begin{center}
    \includegraphics{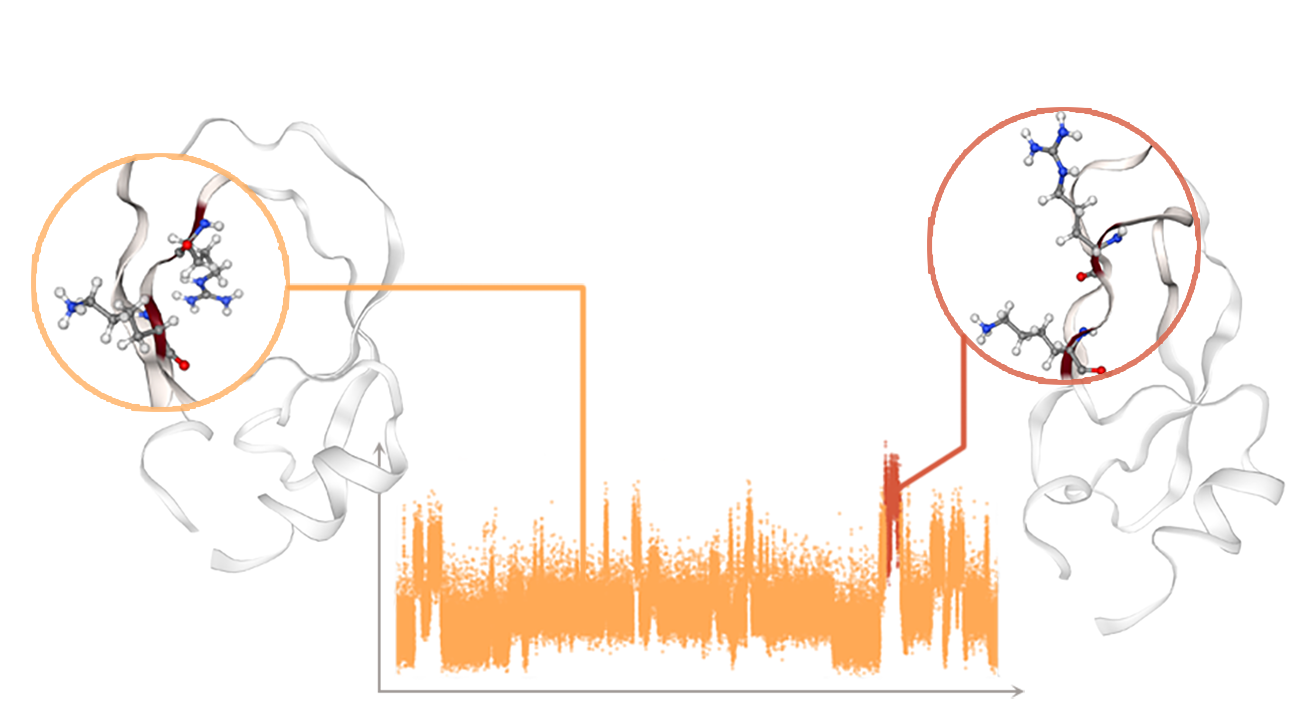}
    \end{center}
    \end{abstract}
    \vspace{1cm}
  \end{@twocolumnfalse}
]
\saythanks

\section{Introduction}
 Atomistic simulations are one of the pillars of modern science. Their impact is felt in areas as diverse as chemistry, material science and biology~\cite{Frenkel2002UnderstandingSimulation}. The fortune of such simulations  have also been  helped  by the great progress made in the last few decades both in algorithms and hardware. This has allowed increasingly more complex systems to be simulated and longer time scales to be explored.  Nowadays, simulations are able to generate a large amount of data and their analysis is becoming more and more challenging. 
 In fact, the analysis of simulation trajectories is a labor intensive endeavor that can take more time than the simulation itself. The purpose of this paper is to cut down the time that researchers spend on this tedious, lengthy and at time frustrating analysis.

What we have in mind is the scenario encountered in many molecular simulations. While the configurational space of a physical system is enormously large, in most cases only a tiny fraction is visited with an appreciable probability. This often leads to a multimodal Boltzmann distribution which reflects the presence of long-lived metastable states. Examples are the initial and final state of a chemical process, the folded and unfolded state of a protein, or the liquid and solid phase of a material. Transitions between those states are rare and occur on a time scale longer than the other characteristic times of the system. Identifying such states and establishing their nature is of course of paramount importance. Several methods for identifying such states have been proposed in the literature, for instance using clustering techniques~\cite{Chang2013PersistentDynamics,Sittel2016RobustProteins} or constructing  disconnectivity graphs~\cite{Krivov2002FreeModels}. When it comes to biomolecular simulations, Markov State Models~\cite{2014AnSimulation} are often used, from which metastable states can be inferred by partitioning microstates~\cite{Deuflhard2000IdentificationChains,Noe2007HierarchicalStates,Martini2017VariationalStates}. Once such states are identified, various machine learning methods can be applied to extract relevant features~\cite{Sultan2014AutomaticSimulations,Brandt2018MachineCoordinates,Fleetwood2020MolecularLearning,Ravindra2019,Bonati2020, Cersonsky2021ImprovingRegression}.

In this work, we propose a data driven method that, starting from a  simulation trajectory, be it biased or unbiased, is  able to characterize the metastable states  of a system in a physically transparent way. To this end, we combine two tools. One is the Variational Approach to Conformation dynamics (VAC) ~\cite{Perez-Hernandez2013,Noe2017CollectiveMethods}. This method allows identifying the slow modes of the system associated with the rare transitions between  metastable states, and ranking them according to their lifetime. This is crucial since the physical notion of metastable states is intimately linked to their lifetime.
We use this criterion to identify the states associated with the slowest transition timescales and  label the configuration explored according to the states they belong to.
The second tool we rely on is the Least Absolute Shrinkage and Selection Operator (LASSO) method \cite{Tibshirani1996RegressionLasso} (see also \cite{Hastie2015StatisticalSparsity,Buhlmann2011StatisticsData} and references therein), a well-known method in machine learning and statistics that performs variable selection via regularization so that both prediction accuracy and interpretability are reached simultaneously. Its role is to extract a handful of critical variables from a large set of physical descriptors. 

We demonstrate the usefulness of this approach on two proteins, a variant of chignolin and the bovine pancreatic trypsin inhibitor (BPTI). In the first case, the data are generated in a biased run, while in the second they are the output of an unbiased 1 ms long trajectory.  In both cases  we are  able to identify clearly their metastable states and describe them in a concise and physically transparent way. In order to present the results in a more illuminating way, we shall make use of the hierarchical structure of the protein free energy landscape and organize the results as a tree of states of decreasing stability in which states with longer time life are analyzed in terms of their shorter lived components. 
    
\section{Methods}\label{sec:methods}

Our method aims to distinguish between different metastable states with models that are physically transparent.
To to this end, we shall employ one of the classical machine learning tools, the LASSO classifier.  Our goal, however, is not merely classification but rather we wish to identify the physical quantities that most characterize the different states. Given an atomic configuration, LASSO takes as input a set  physical descriptors (features) and  outputs the label of the state the configuration  belongs to. We then look for the smallest number of descriptors that still makes the classification accurate. This allows the nature of the different states to be more easily interpreted.
        
\subsection{Choice of physical descriptors}\label{sec:descriptors}
Our source of data is the output of a molecular dynamics simulation, consisting of an atomic trajectory  $\left(\mathbf{R}_{i}\right)_{i = 1}^{N}$, where $i$ runs over the $N$ time steps of the simulation. However, the Cartesian coordinates are not expressive enough to help the researcher to get an  understanding of the states. Because of this, as a first step, we define and evaluate along the trajectory a set of physical descriptors  and use them as inputs for the classifier. With an appropriate choice of descriptors the classification model correctly accounts for the physical symmetries of the system such as invariance to rotations and translations.
We build descriptors out of invariant quantities like distances, angles and dihedral angles. These quantities can be used to construct more elaborate descriptors such as those characterizing bond breaking and forming, protein secondary structures or solvation.

In practice, the number of input descriptors can be very large, of the order of tens of thousands or more. The LASSO algorithm constructs models that use only a handful of such descriptors while being able to distinguish accurately between metastable states. This selection will provide a precious hint toward a physical understanding of the metastable states. 

\subsection{Identifying metastable states}\label{sec:identify}

We now describe the second input needed by the classifier, which is a set of labels corresponding to the metastable state each configuration belongs to. 

To identify automatically the metastable states we use VAC~\cite{Perez-Hernandez2013,Noe2017CollectiveMethods}. In this framework, it is possible to show that the eigenfunctions of the so-called transfer operator~\cite{Prinz2011MarkovValidation} encode the modes that relax more slowly toward equilibrium. To each mode there is also associated a time scale related to the corresponding eigenvalue.

The eigenfunctions of the transfer operator can be estimated using the Time-lagged Independent Component Analysis  (TICA)~\cite{Perez-Hernandez2013,Schwantes2013ImprovementsNTL9} method and can be expressed either as a linear combination of input descriptors or as a nonlinear one using either kernel methods~\cite{Schwantes2015ModelingTrick} or neural networks~\cite{Mardt2018,Chen2019b,Bonati2021DeepSampling}. In the following, we shall use the DeepTICA method~\cite{Bonati2021DeepSampling}, which uses a neural network ansatz for the eigenfunctions and that can deal also with biased simulations. 
The slowest modes of the transfer operator are then used as an analysis tool, for instance to build Markov State Models \cite{Perez-Hernandez2013,2014AnSimulation} or as collective variables $\mathbf{s}(\mathbf{R})$ in enhanced simulations~\cite{McCarty2017c,M.Sultan2017a,Bonati2021DeepSampling}. Here we identify the metastable states starting from the Free Energy Surface (FES) associated with the slowest DeepTICA collective variables (CVs) $\mathbf{s}$. The FES is defined as the logarithm of the marginal distribution of the selected CVs $P(\mathbf{s})$:
\begin{equation}\label{eq:FES_definition}
        \begin{aligned}
        F(\mathbf{s}) &= -\beta^{-1}\ln P(\mathbf{s})\\
        &= -\beta^{-1}\ln \left[\int d\mathbf{R}e^{-\beta U(\mathbf{R})}\delta \left(\mathbf{s}(\mathbf{R}) - \mathbf{s}\right)\right]
        \end{aligned}
    \end{equation}
where $U(\mathbf{R})$ is the potential energy, and $\beta= (k_{{\rm B}}T)^{-1}$  the inverse temperature.
In general the FES is a non-convex function with multiple local minima.
In a rare event scenario these minima correspond to different metastable states and are separated by a high free energy barrier. Hence we identify the metastable states as the minima of the free energy surface projected on the slowest modes.

To locate the metastable basins we use a Gaussian kernel density estimator to approximate the FES from simulation data. In the case of enhanced sampling simulations, the equilibrium probability distribution is recovered via a reweighting procedure~\cite{Valsson2016}. We then identify the different basins of the FES and assign a label to each configuration based on the basin it belongs to. 
Finally, the LASSO classifier is optimized only over those configurations whose free energy difference with respect to the basin's minimum is smaller than a preassigned value (typically 1-2 $k_B T$). In this way we ensure that the samples selected belong to a high-probability region of the phase space and are well separated from the other basins, so that mislabeling is avoided. Details about the algorithms used to construct the FES and identify the local minima are reported in the Supporting Information (Section \ref{app:identifying_minima}). 

\subsection{Interpreting and distinguishing states}\label{sec:classify}
We now move to the core of the present work, i.e.  the construction of a physically interpretable model. To this end we build a function which, given a configuration $\mathbf{R}$, has in input $N_d$ descriptors $d_{1}(\mathbf{R}), \ldots, d_{N_d}(\mathbf{R})$ values and returns in output the  label of the metastable state the  configuration belongs to, see Section~\ref{sec:identify}.  This is a standard supervised classification problem much discussed in the machine learning literature that can be solved in many ways~\cite{Shalev-Shwartz2014UnderstandingLearning}. For the sake of simplicity, we  describe here only the case in which two metastable states are present, labeled here  $-1$ and $+1$. The extension to the general case is simple  and is discussed  in the Supplementary Information (Appendix~\ref{app:multistate_classifier}). The model is then optimized on the values that the $N_d$ descriptors take along the MD trajectory.

Since our scope  is to promote interpretability  we use a  linear model:
\begin{equation}\label{eq:classification_model}
    f(\mathbf{R};\mathbf{w}) = \sum_{j = 1}^{N_d} w_{j}d_{j}(\mathbf{R}).
\end{equation}

With  properly chosen values of the vector   $\mathbf{w} = (w_{1}, \ldots, w_{N_d})$  the function  $f(\mathbf{R};\mathbf{w})$ will be positive or negative according to which label the input $\mathbf{R}$ configuration is assigned.

Having a linear classifier is a first step toward interpretability. However, when one has to deal with thousands of descriptors this is not enough. 
Thus we enforce sparsity using LASSO.
We separate the $N=N_{-}+N_{+}$ atomic configurations into the $N_{-}$ that belong to state $-1$ and the $N_{+}$ that belong to state $+1$. The LASSO method searches for a model that minimizes the empirical error $L$, given by
 \begin{equation}\label{eq:lasso_logistic_regression}
            L(\mathbf{w};\lambda)=  \frac{1}{N}\sum_{\alpha  = \pm1}\sum_{i_{\alpha}=1}^{N_{\alpha}} \log \left[e^{-\alpha f_{\mathbf{w}}(\mathbf{R}_{i_\alpha})} + 1\right] + \lambda\left \Vert \mathbf{w}\right \Vert_{1}.
\end{equation}
where the index $i_\alpha$ runs over the $N_\alpha$ configurations that belong to the $\alpha$ basin.
The first term of the loss function is the standard logistic loss~\cite{Shalev-Shwartz2014UnderstandingLearning}, needed to ensure that the model properly distinguishes the different states, the second enforces the sparsity of $\mathbf{w}$, that is it favors only a handful of components of $\mathbf{w}$ to be different from zero.  The  value of $\lambda$ can be chosen  by evaluating the model classification accuracy on a separated validation set as a function of $\lambda$ and selecting the highest value compatible with high classification accuracy. 
In order to make the choice of $\lambda$ automatic we repeat the training with different values of $\lambda_i$, spaced evenly on a log scale, and use the criterion: 
\begin{equation}
    \lambda = \argmin_{\lambda_i} \left[(1-A(\lambda_i)) * 100 + n_\mathbf{w}(\lambda_i)\right]
    \label{eq:lambda}
\end{equation}
where $A$ is the accuracy over the validation set and $n_\mathbf{w}$ the number of non-zero coefficients at a given $\lambda_i$. This criterion implies that a new feature is included only if the corresponding increase in accuracy is greater than 1\%.

\subsection{Implementation}
We have implemented the methods described above in a Python~\cite{VanRossum2009PythonManual} package designed to be embedded effortlessly in existing MD workflows. The code guarantees maximum flexibility from the point of view of the input features, as the descriptors can be calculated either with  functions provided or externally. This allows applying the method to a wide range of physical, biological and chemical problems.
As the applications presented in this work are related to biological systems, and particularly proteins, we interfaced our code with the Python package \verb_mdtraj_~\cite{McGibbon2015MDTraj:Trajectories}. This allows us to easily calculate angles, dihedrals, relative distances and contacts, as well as more elaborate descriptors. This package, called \verb_stateinterpeter_, is available at: \href{https://github.com/luigibonati/md-stateinterpreter}{https://github.com/luigibonati/md-stateinterpreter}.

\section{Applications}
In this section, we test our method on two different real-case scenarios and explore the free energy landscape of two  proteins. We choose to study two proteins since their  relatively complex behavior  offers an excellent testing ground for our methods. All the results we are about to present are obtained using the \verb_stateinterpeter_  package and can be easily reproduced. 

\subsection{Chignolin}\label{sec:application_chignolin}
We first study the peptide CLN025 (PDB ID 2RVD), which is a variant of the mini-protein chignolin. It is one of the smallest proteins that has a stable folded state. This protein has been extensively studied by brute force molecular dynamics simulations~\cite{Lindorff-Larsen2011HowFold} and enhanced sampling techniques~\cite{Shaffer2016a,Bonati2021DeepSampling}. Since long unbiased simulations are not always available we demonstrate that our method can be used to interpret the more easily obtainable biased simulations. Thus we analyze the trajectories generated in  Ref.~\cite{Bonati2021DeepSampling} in a two-step procedure that started from a multicanonical-type simulation.  In  Ref.~\cite{Bonati2021DeepSampling}, the slowest modes of this biased simulation were evaluated along with the corresponding FES.
However, it took time and effort to arrive at a physical understanding of the four different states thus found.
  
We shall show instead that the tools developed here allow repeating the analysis of Ref.~\cite{Bonati2021DeepSampling} in a matter of seconds in an automatic and effortless way. In addition, our analysis reveals that in the Chignolin case and even more clearly in the following example of BPTI that the free energy has a hierarchical structure in  which slow modes can be decomposed into faster ones.

\begin{figure}
    \centering
    \includegraphics{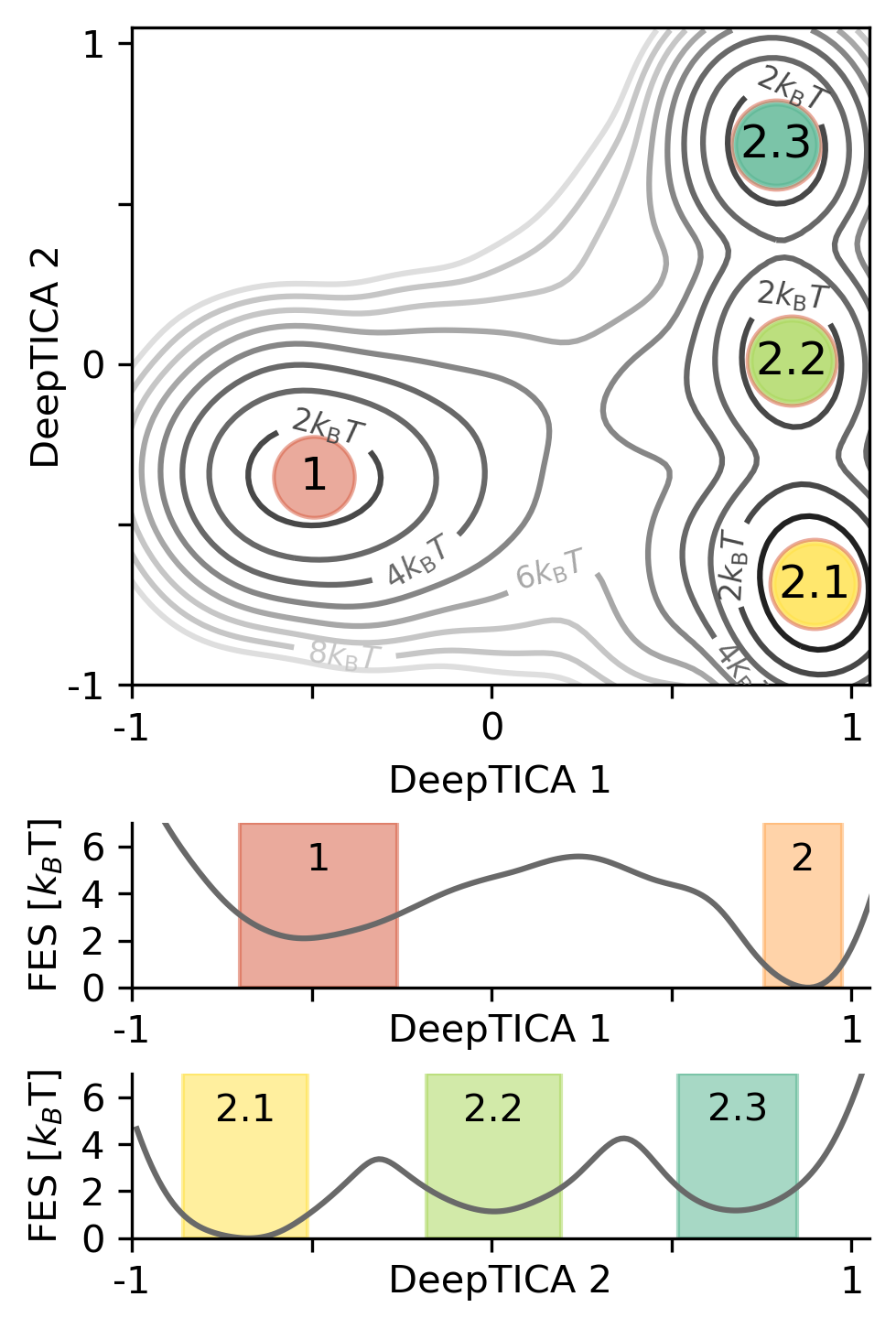}
    \caption{
    (top) Isolines of the free energy surface as a function of the slowest approximated eigenfunctions Deep-TICA 1 and 2. Each minimum is labeled with the name of the metastable state. (center) Free energy projected along DeepTICA 1. (bottom) Free energy projected along DeepTICA 2, constructed using the data belonging to state 2 (identified from DeepTICA 1).
    In both the central and bottom plot the highlighted regions around the minima correspond to the windows in which the samples are selected for the classification.
    }
    \label{fig:chignolin_states}
\end{figure}

The hierarchical structure of the eigenfunctions of the transfer operator becomes apparent if we look at the free energy surface computed as a function of the two leading functions, see Fig.~\ref{fig:chignolin_states}. The first one, DeepTICA 1 is associated with the slowest transition between states 1 and 2, while the latter describes the transformation between shorter-lived metastable states nested inside state 2.

To make use of this hierarchical structure, we start by projecting the free energy along the approximated eigenfunction DeepTICA 1 and characterizing the two states thus identified.  In our analysis we first use as descriptors the contact functions between all the 318 donor-acceptor pairs in the protein that can form a hydrogen bond. These are defined as 

\begin{equation}\label{eq:chignolin_descriptors_contacts}
    \phi(r) = \frac{1 + (r/r_{0})^{6}}{1 + (r/r_{0})^{12}}.
\end{equation}
Here, $r$ is the donor-acceptor distance and $r_0=3.5$\AA~is the typical distance between two heavy atoms involved in a H-bond. Hence, Eq.~\eqref{eq:chignolin_descriptors_contacts} reports on whether an H-bond is formed.  

In the following, we primarily discuss the descriptors selected by the sparse classifier and how they can help to characterize states. In the Supplementary Information (SI) we illustrate the behavior of the models in terms of accuracy and complexity as the regularization parameter changes (Fig.~\ref{fig:supporting_chignolin_accuracy_complexity}) and the histograms of the selected descriptors in different states (Fig.~\ref{fig:supporting_chignolin_histogram}).

\begin{figure}
    \centering
    \includegraphics[width=\linewidth]{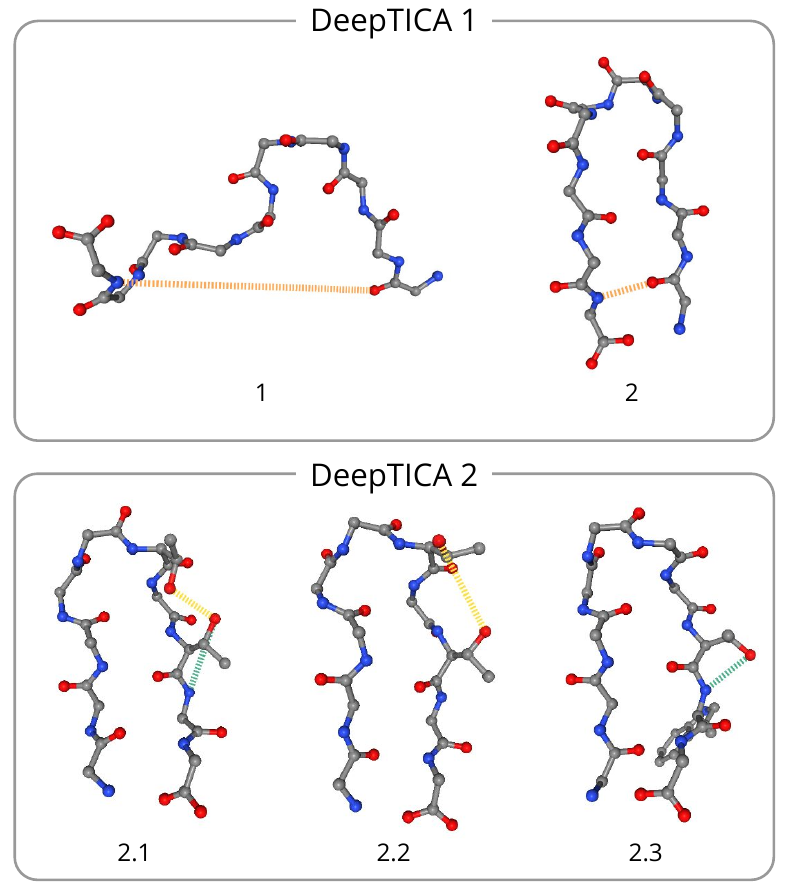}
    \caption{
    Snapshots of the conformations of the proteins in each state, for the DeepTICA 1 (top) and DeepTICA 2 (bottom) variables. We also report the H-bonds selected by LASSO for each state. For DeepTICA 2 the features relevant to the two binary classifiers are displayed: in yellow the one related to state 2.1 vs 2.2 and in green the one associated with states 2.1 vs 2.3. Only the backbone of the protein is shown, but for the residues in which a feature of the sidechain is selected. The protein snapshots are obtained with the NGLview iPython widget ~\cite{Nguyen2018NGLviewinteractiveNotebooks}.
    }
    \label{fig:chignolin_snapshots}
\end{figure}

Out of 318 possible acceptor-donor contacts, the sparse estimator (Eq.~\eqref{eq:lasso_logistic_regression}) finds that a single contact is sufficient to distinguish between the states. They are indeed the unfolded (1) and folded (2) state, as can be easily determined also by a visual inspection of the trajectory. This is hardly surprising since in this system the slowest mode is associated with the folding-unfolding transition. Nevertheless, this is a physically transparent result since the bond between the terminal residues closes the $\beta$ hairpin structure of the folded state and thus it is able to distinguish clearly between them (Fig.~\ref{fig:chignolin_snapshots}).

\begin{table*}[h!]
\centering

\begin{tabular}{|ccc|}
\hline
\multicolumn{3}{|c|}{\textbf{DeepTICA 1}}                                                                                                                                                                                 \\ \hline
\multicolumn{1}{|c|}{States}                      & \multicolumn{1}{c|}{\begin{tabular}[c]{@{}c@{}}Distances (99\%)\end{tabular}}  & \begin{tabular}[c]{@{}c@{}}Dihedrals (99\%)\end{tabular}  \\ \hline
\multicolumn{1}{|c|}{1 vs 2}                      & \multicolumn{1}{c|}{TYR1-O -- TYR10-N}                                                       & ASP3-$\chi_1$                                      \\ \hline
\multicolumn{3}{|c|}{\textbf{DeepTICA 2}}                                                                                                                                                                                 \\ \hline
\multicolumn{1}{|c|}{States}                      & \multicolumn{1}{c|}{\begin{tabular}[c]{@{}c@{}}Distances (96\%)\end{tabular}}   & \begin{tabular}[c]{@{}c@{}}Dihedrals (96\%)\end{tabular}  \\ \hline
\multicolumn{1}{|c|}{\multirow{2}{*}{2.1 vs 2.2}} & \multicolumn{1}{c|}{\multirow{2}{*}{THR6-Os -- THR8-Os}}                                     & THR6-$\chi_1$                                                          \\ \cline{3-3} 
\multicolumn{1}{|c|}{}                            & \multicolumn{1}{c|}{}                                                                        & THR8-$\chi_1$                                                          \\ \hline
\multicolumn{1}{|c|}{States}                      & \multicolumn{1}{c|}{\begin{tabular}[c]{@{}c@{}}Distances (100\%)\end{tabular}} & \begin{tabular}[c]{@{}c@{}}Dihedrals (100\%)\end{tabular} \\ \hline
\multicolumn{1}{|l|}{2.1 vs 2.3}                  & \multicolumn{1}{c|}{THR8-Os -- TRP9-N}                                                       & THR8-$\chi_1$                                                          \\ \hline
\end{tabular}
    \caption{
    Descriptors selected by the LASSO estimator for the chignolin system. Each row contains the result of the classifiers for the states identified by a given DeepTICA CV, either using the H-bond contact functions or dihedral angles. The percentages in parenthesis indicate the classification accuracy. The ``s'' suffix in the H-bond contacts identifies side chain atoms.
    }
    \label{tab:chignolin_descriptors}
\end{table*}

However, on the shorter time scale of the second TICA eigenvalue, state 2 (which corresponds to a set of folded states) reveals a finer structure. This can be seen by selecting only the configurations that belong to state 2 and projecting them on DeepTICA2 (see Fig.~\ref{fig:chignolin_states}).  
To deal with the case of more than two states of course one can resort to a multi-class classifier, e.g. composed by N binary classifiers, each discriminating a given state from all the others (Section~\ref{app:multistate_classifier}). However, we find closer to our idea of interpretation discriminating two states at the time rather than comparing all the states  together. For this reason we decided to inspect the differences between the most stable state (2.1) and the others.
As can be seen in the lower panel of Fig.~\ref{fig:chignolin_snapshots}, the relevant features are connected to the sidechains of THR6 and THR8 (see also Table~\ref{tab:chignolin_descriptors} and Fig.~\ref{fig:supporting_chignolin_histogram})). Notably, the most stable state is characterized by the bonding between these two sidechains. This result is coherent with our previous analysis~\cite{Bonati2021DeepSampling} and in line with the finding that the interaction between THR6 and THR8 plays a crucial role in the stabilization of the folded state of chignolin~\cite{Maruyama2018AnalysisChignolin}. 
This result highlights the importance of being able to investigate a large set of descriptors, including also the often neglected sidechains, which in this case play a relevant role.

It is instructive to repeat this analysis using a different set of descriptors. As an alternative to the H-bond descriptors, we use the dihedral angles related both to the conformational transformations of the backbone (Ramachandran angles $\phi$ and $\psi$) and to that of the sidechains ($\chi_1$ and $\chi_2$). To take properly into account the periodic nature of these variables we use as descriptors the sine and cosine of the angles, this leads to a total of 68 variables.
 
The angle-based LASSO results, summarized in Table~\ref{tab:chignolin_descriptors}, are consistent with that based on the H-bonds, as the angles distinguishing the states are related to the residues that were identified as crucial in the H-bond-based analysis. The only apparent exception is the ASP3-$\chi_1$ angle that discriminates state 1 (unfolded) from 2 (folded). The fact that this side chain angle is selected  might be surprising. However, one must not forget that the output of this analysis are the descriptors that discriminate the most, and do not provide any insight as to the why. In fact,  ASP3-$\chi_1$  is selected because, in the folded state, it is rigidly locked at 180 degrees by the hydrogen bond formed between residue 3 and residue 8. It is therefore a byproduct of the folding/unfolding process. For this reason, if we do not know already which descriptors are the most suitable, it is best to characterize the same process with different sets of descriptors: doing so provides a concise representation from multiple perspectives that can aid understanding. Indeed, we recall that our purpose is exactly to help focusing on the physically interesting regions.

Before leaving this section a few more comments are in order. One could have continued the analysis and resolved each of the 3 substates discussed earlier into shorter lived modes. However, since the barriers between modes are of the order of $k_BT$, the exercise of identifying these fast modes would have become arbitrary and below the resolution  of our analysis. We would also like to recall that we have started from data obtained in a multithermal sampling. Thus, what we analyze here are the slow modes of such simulation which, in principle, could differ from those of an unbiased one. However, as far as the modes described here are concerned we find the same results if we repeat the analysis on the unbiased data. Indeed, as reported in the SI (Section \ref{app:chignolin_unbiased}) both the structure of the metastable states and the descriptors selected are practically the same. 

\subsection{BPTI}\label{sec:application_BPTI}

\begin{figure*}
    \centering
    \includegraphics[width=\textwidth]{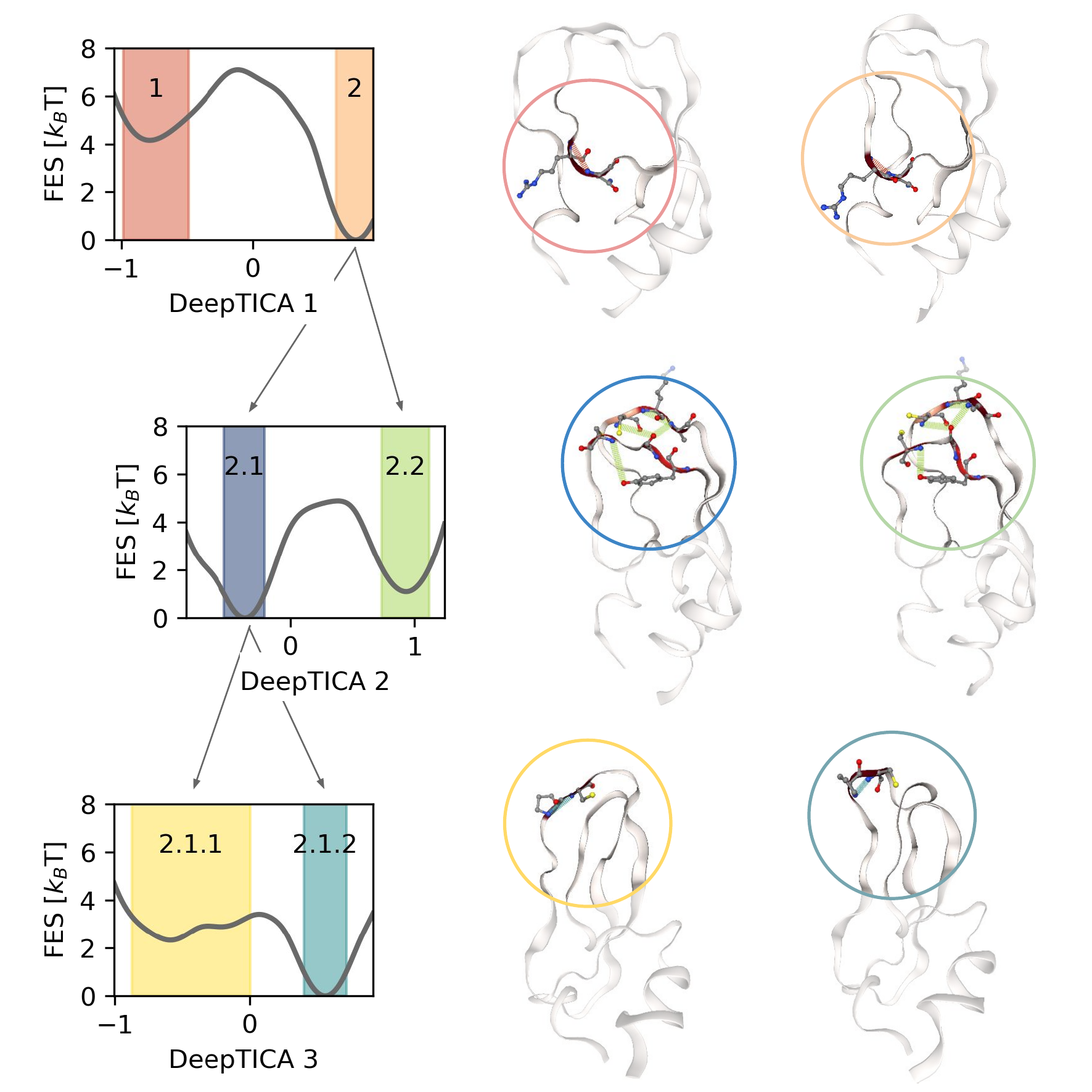}
    \caption{(left) Hierarchical presentation of the free energy surfaces projected along with the slowest DeepTICA CVs. The free energy profiles associated with the eigenfunctions faster than DeepTICA 1 have been constructed using a slice of the data taken from the basins of the previous CV. Only the projections that have more than 1 state are shown.
    (right) Snapshots of BPTI in which the relevant H-bond contacts selected by LASSO are reported. The circle, colored according to the state, focuses the attention on the region containing the relevant features.
    }
    \label{fig:bpti_hierarchical}
\end{figure*}

The second example studied is the 58-residue long BPTI protein (PDB ID 5PTI). The data that we base our analysis on come from a 1 ms long molecular dynamics simulation performed using the specialized ANTON~\cite{Shaw2010Atomic-levelProteins} computer. In the original work, a complex clustering procedure based on autocorrelation functions and on the introduction of ad-hoc functions that measure the residues' significance was used to identify and understand the different metastable states. Some analysis of the same run has been performed using the sparse TICA method~\cite{McGibbon2017IdentificationDynamics}. 

Unfortunately, ANTON's trajectory is not long enough to observe the unfolding  transition and even within the set of folded states one state is visited only once. However, our approach is powerful enough to allow a detailed analysis of the different metastable states and their hierarchical structure even in the presence of an imperfect sampling.

Following the workflow described in Section~\ref{sec:methods}, we first apply Deep-TICA to the ANTON's trajectory and identify the slowest dynamical modes. We use as descriptors the 10271 distances between all the pairs of O and N atoms that can form hydrogen bonds. A fully connected neural network parameterized by four layers and [1024,256,64,3] nodes per layer is used to express the three slowest eigenfunctions of the transfer operator, using a lag-time equal to 1 $\mu s$.

To take full advantage of the hierarchical structure we proceed as in the chignolin example. That is, we look at the  FES projected along the eigenfunctions ordered according to the time scales of the transitions. The data in the different minima  of the longer-lived metastable states are projected into the faster TICA eigenfunctions.  This allows us to organize the states hierarchically (see Fig.~\ref{fig:bpti_hierarchical}) according to their lifetime.  From the LASSO point of view, this helps us query the data in a more structured manner, and focus our attention on the  states that more usefully need to be compared.

\begin{figure}
    \centering
    \includegraphics[width=0.95\linewidth]{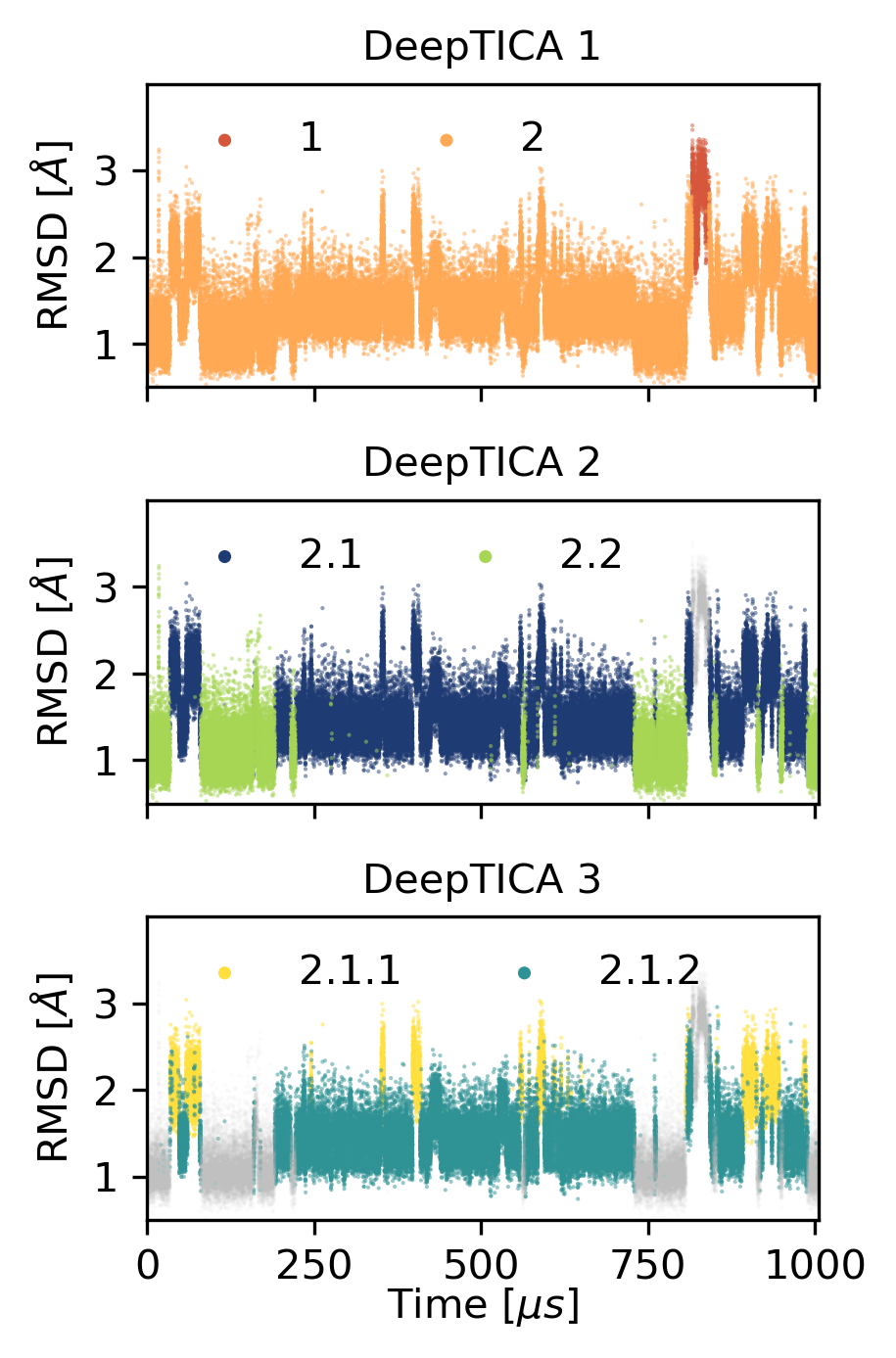}
    \caption{
    Alpha-carbon root-mean-square-deviation with respect to the crystal structure of BPTI. The same data is displayed in all three plots, but each time it is colored according to the states identified by the corresponding DeepTICA CV. Points colored in grey with high transparency belong to other basins of the hierarchical landscape.
    }
    \label{fig:bpti_rmsd}
\end{figure}

On the time scale of the largest Deep-TICA eigenvalue, the eigenfunction DeepTICA 1 distinguishes between state 1  and a second state 2. Since a transition to state 1 has been seen only once along the trajectory (Fig.~\ref{fig:bpti_rmsd}a), it is not surprising that DeepTICA 1 identifies the transition to and from state 1 as a slow mode.  Of course, the difference in free energy between state 1 and state 2 is not to be trusted due to insufficient sampling. However, this fact has no consequence on the analysis that follows, since  we use only configurations close to the minima.

When projected on DeepTICA 2, state 2 as identified by DeepTICA 1 decomposes into two faster modes 2.1 and 2.2. Mode 2.2 is the conformation closest to the crystallographic one, as measured by  the RMSD  (Fig.\ref{fig:bpti_rmsd}), but it is not the most stable structure for this force-field.
Continuing the analysis and using DeepTICA 3, state 2.1 can be further decomposed into two states  2.1.1  and 2.1.2. However here we are at the limits of the resolution of our method and these two states are not very stable, as the free energy barrier in going back to the one with the lowest free energy (2.1.2) is rather small.
It is already interesting to note that our identification of the metastable states agrees with the clustering analysis performed in Ref.~\cite{Shaw2010Atomic-levelProteins}, as easily checked by comparing their results with our Fig.~\ref{fig:bpti_rmsd}. 

As in the previous example, we compare the performances of  two sets of descriptors: the contact functions between possible H-bonded   pairs    and the dihedral angles $\phi,\psi,\chi_1$ and $\chi_2$, to find out which set discriminates best the states.  Out of the 10271 possible H-bonded  pairs, we use as descriptors only those that have at least once formed an H-bond, as measured by the contact function in Eq.~\eqref{eq:chignolin_descriptors_contacts} being greater than 0.5. This leads to the selection of 927 descriptors. When we use the angular variables, given the fact that we use the sine and cosine of the dihedrals we end up with a total of 392 descriptors.

 The results for these two LASSO analyses are shown in Table~\ref{tab:bpti_lasso}. 
Both analysis achieve very high accuracy and identify as relevant the same residues. In particular, for the  DeepTICA 1 related transition (state 1 vs 2) we find that the residue ARG42 plays an important role. To discriminate between the states associated with DeepTICA 2 (2.1 and 2.2) LASSO outputs a few contacts between residues 14-15-16 and 35-36-38, while for the dihedrals the $\psi$ angle of LYS15.
Finally, for the states associated with DeepTICA 3 (2.1.1 vs 2.1.2), the important descriptors are in both cases associated with the residue PRO13. This  allow us to look at the protein by focusing our view on the most affected regions, as shown Fig.~\ref{fig:bpti_hierarchical}. 

\begin{table*}
    \centering
\begin{tabular}{|ccc|}
\hline
\multicolumn{3}{|c|}{\textbf{DeepTICA 1}}                                                                                     \\ \hline
\multicolumn{1}{|c|}{States}                      & \multicolumn{1}{c|}{H-bond contacts (99\%)  }            & Dihedrals (99\%)                    \\ \hline
\multicolumn{1}{|c|}{\textbf{1 vs 2}}                      & \multicolumn{1}{c|}{ARG42-N -- ASN43-N}   & $\phi$ ARG42                  \\ \hline
\multicolumn{3}{|c|}{\textbf{DeepTICA 2}}                                                                                     \\ \hline
\multicolumn{1}{|c|}{States}                      & \multicolumn{1}{c|}{H-bond contacts (98\%)}            & Dihedrals (96\%)                      \\ \hline
\multicolumn{1}{|c|}{\multirow{4}{*}{\textbf{2.1 vs 2.2}}} & \multicolumn{1}{c|}{TYR35-Os -- CYS38-N} & \multirow{4}{*}{$\psi$ LYS15} \\ \cline{2-2}
\multicolumn{1}{|c|}{}                            & \multicolumn{1}{c|}{CYS14-N -- GLY36-O}   &                               \\ \cline{2-2}
\multicolumn{1}{|c|}{}                            & \multicolumn{1}{c|}{LYS15-N -- ALA16-N}   &                               \\ \cline{2-2}
\multicolumn{1}{|c|}{}                            & \multicolumn{1}{c|}{ALA16-N -- GLY36-O}   &                               \\ \hline
\multicolumn{3}{|c|}{\textbf{DeepTICA 3}}                                                                                     \\ \hline
\multicolumn{1}{|c|}{States}                      & \multicolumn{1}{c|}{H-bond contacts (97\%)}            & Dihedrals (98\%)                       \\ \hline
\multicolumn{1}{|c|}{\textbf{2.1.1 vs 2.1.2}}              & \multicolumn{1}{c|}{PRO13-N -- CYS14-N}   & $\psi$ PRO13                  \\ \hline
\end{tabular}

    \caption{Descriptors selected by the LASSO estimator for the BPTI system. Each row contains the result of the classifiers for the states identified by a given DeepTICA CV, either using the H-bond contact functions or dihedral angles. The percentages in parenthesis indicate the classification accuracy. The ``s'' suffix in the H-bond contacts identifies side chain atoms.
    }
    \label{tab:bpti_lasso}
\end{table*}

Regarding state 1, it is interesting to note that the $\phi$ angle of the ARG42 residue is the same variable found by applying the sparse TICA method in Ref.~\cite{McGibbon2017IdentificationDynamics}. This angle flips when the system makes a transition to state 1, although in Ref.~\cite{McGibbon2017IdentificationDynamics} it was already noted that this is actually a byproduct of a large-scale structural change involving the opening of the protein core. Indeed, if we visualize the relevant residues in a snapshot of state 1 compared to the crystal structure (Fig. ~\ref{fig:supporting_bpti_snapshots}) we observe that these residues experience significant changes leading to the opening of the upper region of protein.
This is not surprising, since our approach, just like the one used in Ref.~\cite{McGibbon2017IdentificationDynamics}), looks for the best \textit{linear} combination of input features capable of performing the task, which here is to distinguish between states.
One could repeat the analysis by employing different set of descriptors which are capable of measuring different physical properties. In the SI we report such analysis using as descriptors the contributions of each residue to the solvent accessible surface area (SASA). In this way, we find that the residues ARG39 and LYS41 undergo a significant change in their SASA contributions (see Fig.~\ref{fig:supporting_bpti_sasa}). Remarkably, not only these residues are very close spatially to the ones selected by the previous analysis, but also in Ref.~\cite{Shaw2010Atomic-levelProteins} the same ARG39, LYS41 were identified among the residues most characterizing this state. 

\section{Conclusions}
In this work, we presented a machine learning-based method to gain physical insights into the properties of metastable states of complex systems. LASSO, the machine learning algorithm at the core of our method, is highly scalable and rests on solid theoretical properties~\cite{Hastie2015StatisticalSparsity,Buhlmann2011StatisticsData}.
Another pillar of our approach is an automated procedure that identifies metastable states from the free energy basins of longer lived slow modes coordinates. To characterize such states we make use of a class of easily interpretable models that return an accurate description of the system in terms of a small number of physical quantities. Using sets of descriptors that are sensitive to different physical properties, the nature of the metastable states can be assessed from multiple perspectives. This workflow is naturally apt to interpret the results of both biased and unbiased simulations.

Although the examples given here are protein-related, this workflow is very general, and we believe it can be useful for analyzing very different simulations, from chemical reactions to material science problems.

\section*{Acknowledgements}
We thank D. E. Shaw Research for sharing their BPTI trajectory.
L.B. thanks Dr. Narjes Ansari for useful discussions and for guidance in the analysis of BPTI simulation. P.N. was in part supported by the ELISE grant (GA no. 951847).

\clearpage

{
\footnotesize
\printbibliography
}

\clearpage

\renewcommand{\thesection}{S\arabic{section}}
\renewcommand{\thesubsection}{S\arabic{subsection}}
\renewcommand{\thefigure}{S\arabic{figure}}
\renewcommand{\theequation}{S\arabic{equation}}
\renewcommand{\thetable}{S\arabic{table}}
\renewcommand{\thepage}{SI-\arabic{page}}
\setcounter{page}{1}  
\setcounter{figure}{0}  
\setcounter{equation}{0} 
\setcounter{section}{0}

\section*{SUPPORTING INFORMATION}
\section{Identifying local minima of the free energy surface}\label{app:identifying_minima}

To identify the metastable basins we approximate the free energy surface with a standard Gaussian kernel density estimator~\cite{Scott1992MultivariateVisualization.}. Next, we seek the local minima of the approximated FES. Every local minima will identify a basin and hence a metastable state. The identification of the local minima of the approximated FES is performed by standard descent algorithms. Descent algorithms are iterative methods which require an initial point to start the iterations, and converge to a near local minimum. We use a basic strategy to retrieve local minima. We initialize multiple BFGS solvers~\cite{Nocedal2006NumericalOptimization} at different initial points, and we let them converge. Finally we recover the set of unique local minima found. The initial points can be selected either by a grid or random sampling of the region of interest. A grid sampling is very accurate, but the number of points grows exponentially fast with the dimension of the CVs space. For the random sampling case, a predefined number of points is sampled from the region of interest according to a given probability distribution (e.g. uniform). Our empirical tests shows that for the problems analyzed, both initialization schemes works very well. If physically not relevant shallow local minima are present in the FES one can exclude them annealing the iterations of the descent algorithm by adding noise~\cite{Wales1997GlobalAtoms}. Once the local minima have been found, we partition the CV space with a Voronoi tesselation centered at the local minima. With the CV space partitioned as such, labeling a configuration requires only to identify in which of the Voronoi polyhedrons it belongs.

If we are only interested in a single CV, a simpler approach is possible. By spltiting the 1D domain of the FES in segments whose extrema correspond to local {\it maxima} of the FES, we obtain a partitioning of the (1D) CV space in which each segment contains a single metastable basin of the FES.

\section{Interpreting three or more metastable states}\label{app:multistate_classifier}
In this section we describe how the techniques presented in Section~\ref{sec:classify} can be extended to give a joint interpretation on a number $N_{{\rm s}} > 2$ of metastable states. In this case, we adopt the ``one-versus-rest'' scheme to distinguish the states. Within this scheme, $N_{{\rm s}}$ disjoint problems of the form of Eq.~\eqref{eq:lasso_logistic_regression} are solved. In the $i$-th problem we split the configurations into the $N_{+1}$ belonging to the state $i$ and the $N_{-1}$ not belonging to $i$. In this way, the solutions $\mathbf{w}_{1},\ldots,\mathbf{w}_{N_{{\rm s}}}$ are associated with the models $f(\mathbf{R}; \mathbf{w}_{i})$ whose sign is positive when we can identify the configuration $\mathbf{R}$ with the state $i$ and negative otherwise. The ``one-versus-rest'' scheme was specifically selected as it preservers the overall intepretability of the results. Indeed, the descriptors selected in the ``one-versus-rest'' scheme for the $i$-th model are the ones needed to distinguish a configuration in the $i$-th state from any other state. More advanced classification methods such as the ones based on the cross-entropy loss do not retain this easy interpretation of the results.

\section{Numerical details}
For the KDE approximation of the FES we have used a bandwidth $\sigma = 0.075$ in the chignolin example presented in Sec.~\ref{sec:application_chignolin} and $\sigma = 0.1$ for the BPTI example in Sec.~\ref{sec:application_BPTI}. In both cases, the collective variables were rescaled to fit in the interval $[-1, 1]$. The training of the LASSO model was performed with 10000 sample configurations for each metastable state.

\section{Chignolin, supplementary results}\label{app:chignolin}

\begin{figure}[htp]
\centering
\subfloat{%
  \includegraphics[clip,width=0.95\columnwidth]{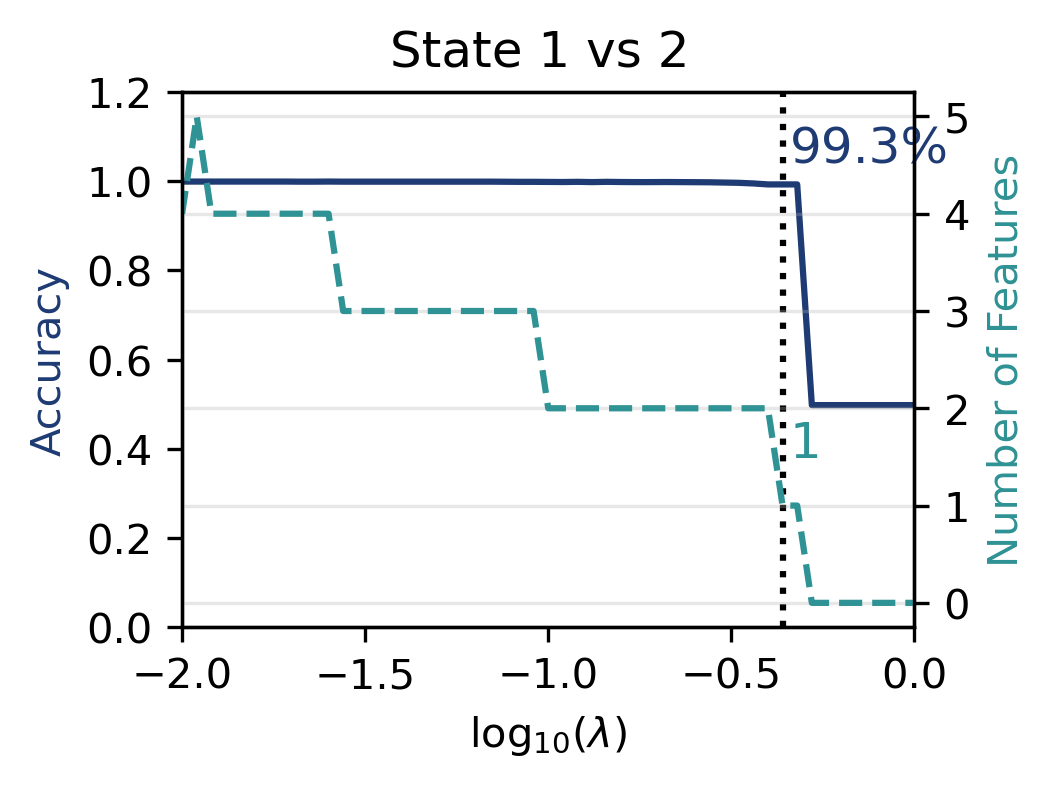}%
}

\subfloat{%
  \includegraphics[clip,width=0.95\columnwidth]{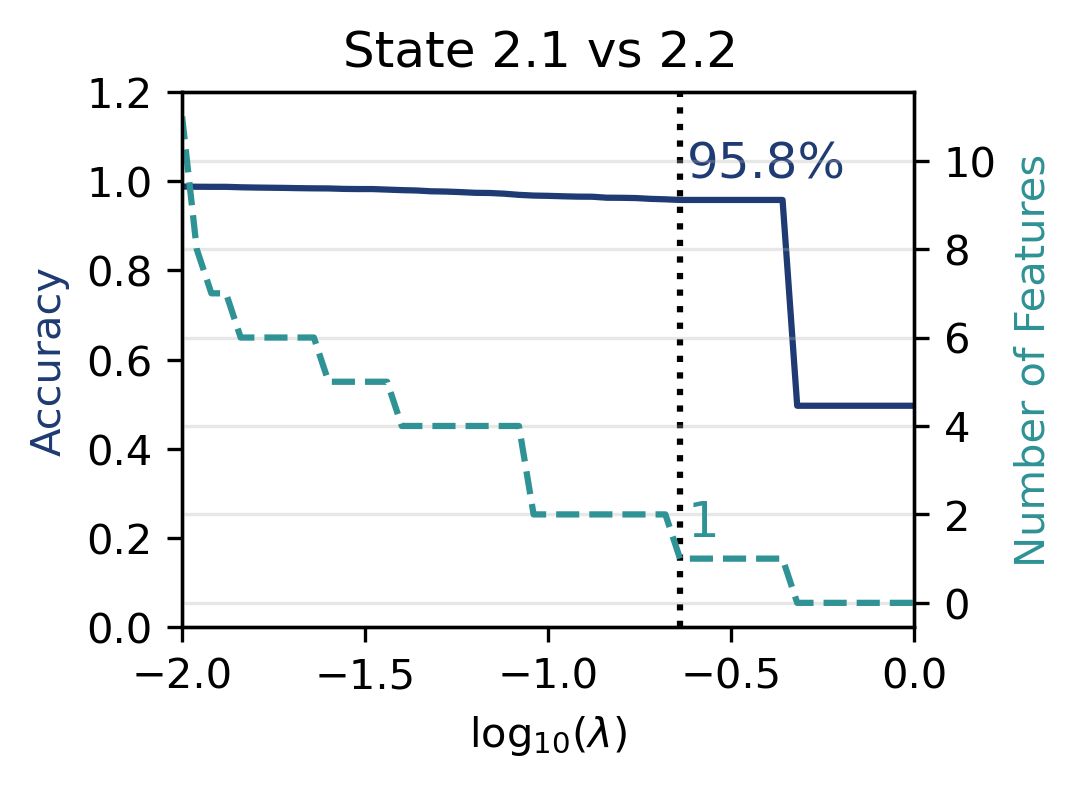}%
}

\subfloat{%
  \includegraphics[clip,width=0.95\columnwidth]{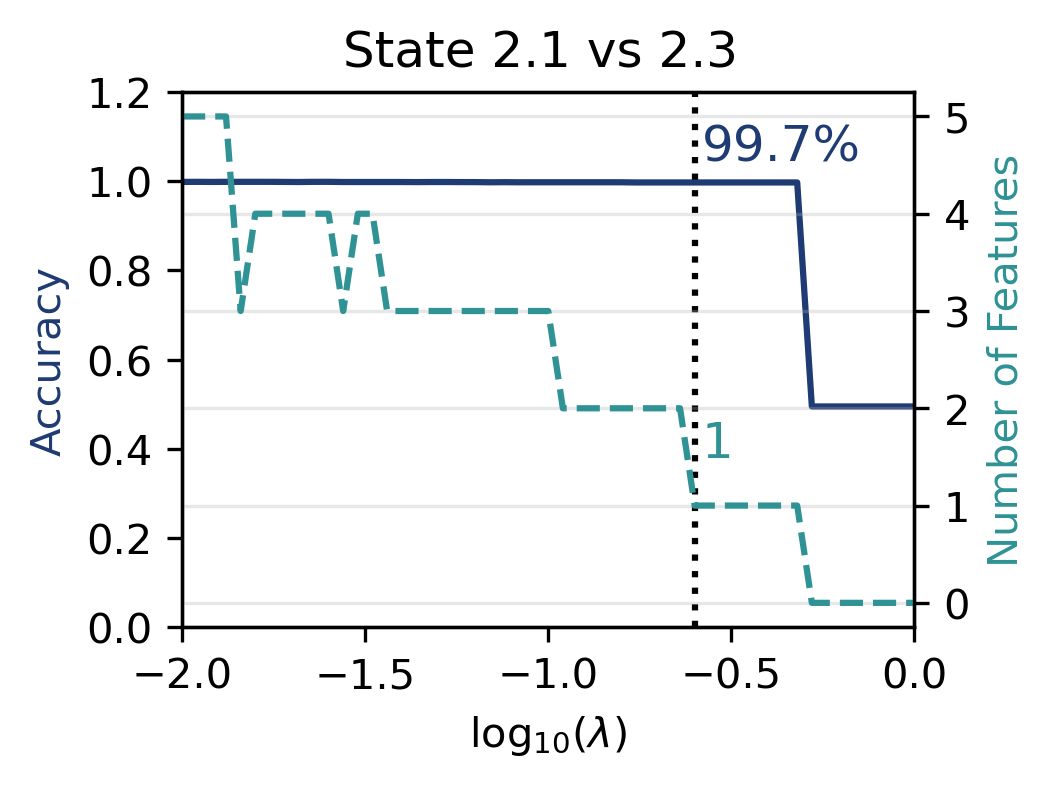}%
}

\caption{Chignolin. Accuracy and number of features as a function of the magnitude of the LASSO regularization. The dotted vertical line represent the value selected by the criterion of Eq.~\ref{eq:lambda}.}
\label{fig:supporting_chignolin_accuracy_complexity}
\end{figure}

\begin{figure}[htp]
\centering

\subfloat{%
  \includegraphics[clip,width=\columnwidth]{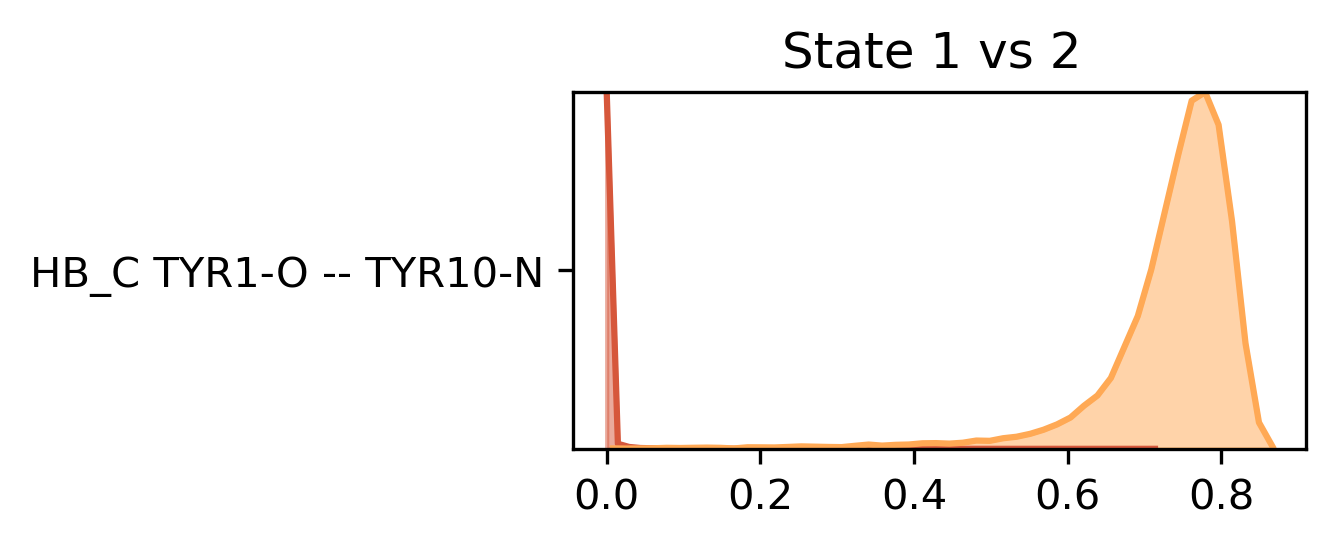}%
}

\subfloat{%
  \includegraphics[clip,width=\columnwidth]{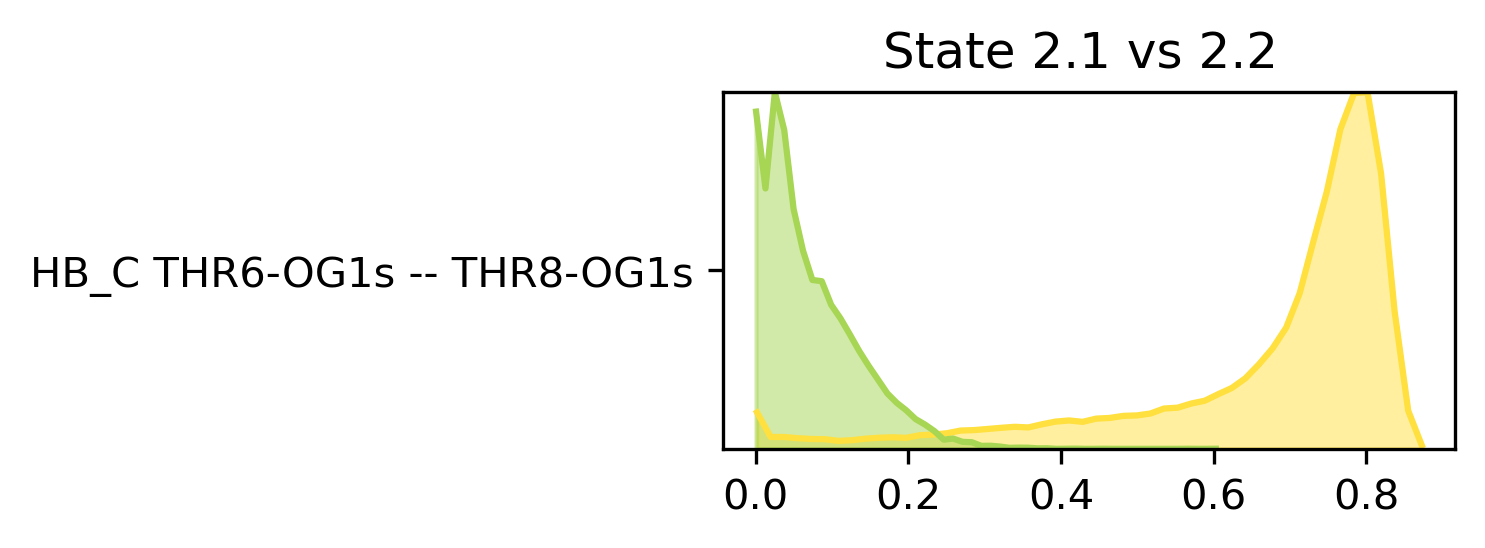}%
}

\subfloat{%
  \includegraphics[clip,width=\columnwidth]{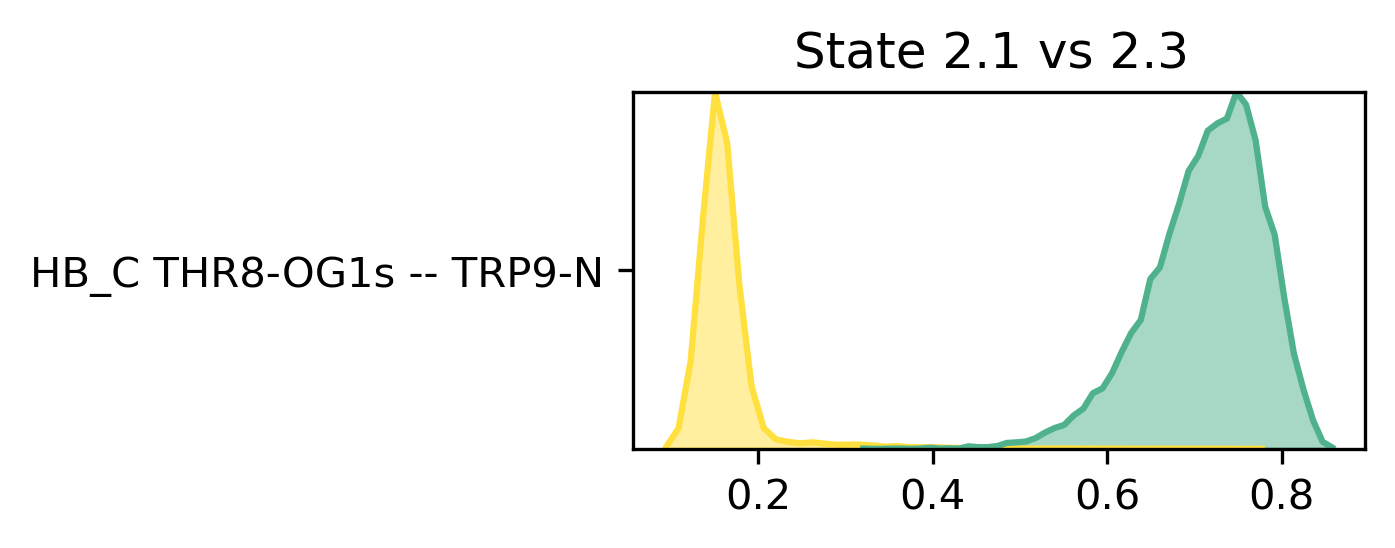}%
}

\caption{Chignolin. Histograms of the the selected features (h-bond contacts) selected by the three classifiers. Colors denote state membership according to Fig.~\ref{fig:chignolin_states}. Note that the histograms are normalized within each state.}
\label{fig:supporting_chignolin_histogram}
\end{figure}

Fig.~\ref{fig:supporting_chignolin_accuracy_complexity} reports the plots of the accuracy and complexity (number of features) for the LASSO classifier as a function of the regularization strength. In fig~\ref{fig:supporting_chignolin_histogram} we plot the histograms of the descriptors selected by LASSO for the different states, to highlight the formation/breaking of the associated hydrogen bonds.

\section{Chignolin, analysis on unbiased data}\label{app:chignolin_unbiased}

Here we repeat the analysis on the Chignolin protein, this time using the data from the molecular dynamics simulation of ref.~\cite{Lindorff-Larsen2011HowFold} rather than from the biased simulation of ref.~\cite{Bonati2021DeepSampling}. The DeepTICA CVs are extracted using a descriptors set composed of all the contacts functions between atoms that can form H-bonds, filtered to keep only the ones that have formed a contact at least in one trajectory snapshot. This gives us 318 input descriptors. Then a neural-network with [318,64,64,3] neurons per layer is used to approximate the eigenfunctions of the transfer operator, as in~\cite{Bonati2021DeepSampling}. The lag-time used is 2 ns. 

\begin{figure*}[htp]
\centering
\subfloat{%
  \includegraphics[clip,width=0.95\columnwidth]{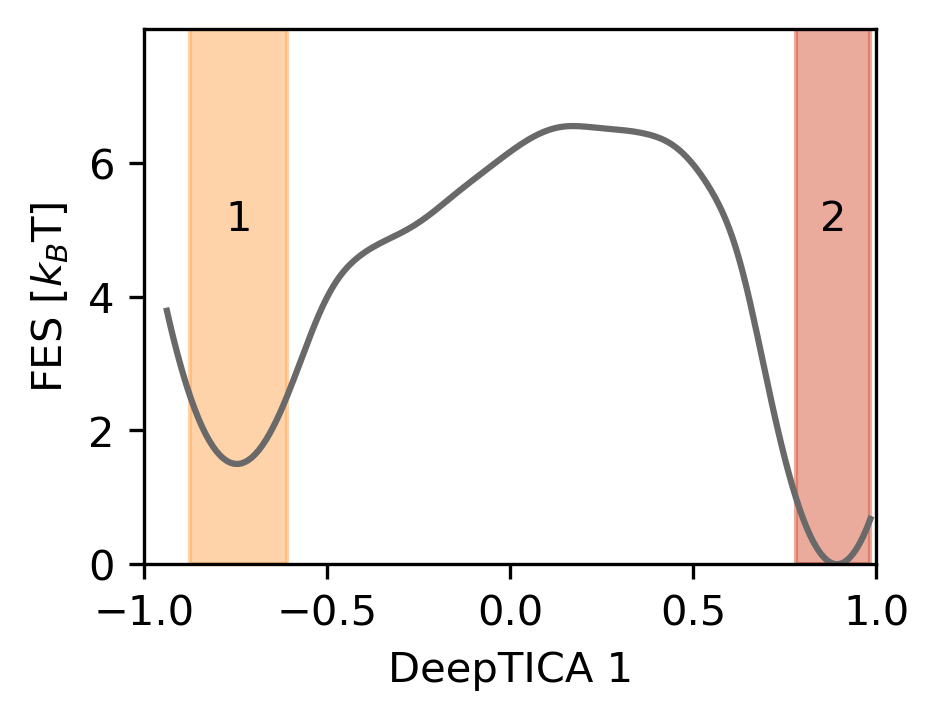}%
}
\subfloat{%
  \includegraphics[clip,width=0.95\columnwidth]{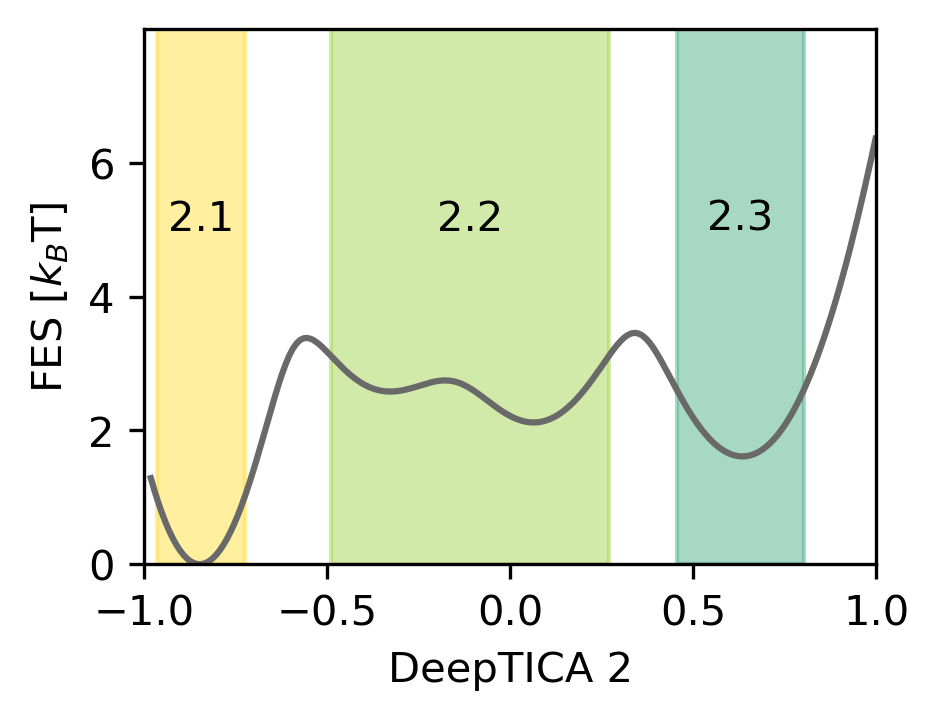}%
}

\caption{Chignolin, unbiased. Hierarchical free energy surfaces as a function of the DeepTICA CVs extracted from the molecular dynamics simulation of ref.~\cite{Lindorff-Larsen2011HowFold}. Colored regions denote the states identified For DeepTICA 2, only the points belonging to basin 2 are used to construct the FES.}
\label{fig:supporting_chignolin_unbiased_fes}
\end{figure*}

Once we have extracted the new CVs, we can identify the metastable states and characterize them with LASSO. The free energy surfaces along the DeepTICA CVs are reported in Fig.~\ref{fig:supporting_chignolin_unbiased_fes}. The structure of the metastable basins is very similar to the one obtained from the biased simulation (Fig.~\ref{fig:chignolin_states}). In particular, DeepTICA 1 describes the transition between two states while DeepTICA 2 describes a faster transition between three states.  Then we identify the most distinguishing features with LASSO, which are reported in Table~\ref{tab:supporting_chignolin_unbiased_descriptors}. Remarkably, the H-bonds contacts selected are the same as Table~\ref{tab:chignolin_descriptors}. Also for the torsional angles-based analysis we find the same descriptors, but for the exchange of $\chi_1$ of THR6 and THR8 in the case of states 2.1 vs 2.3. 

As detailed in the main text, this agreement between the description of slow modes of biased and unbiased simulation was not a foregone conclusion. What is interesting is that with the approach presented in this paper we are able to readily compare states identified by different "cryptic" variables such as neural networks, and show that they identify transitions between the same states. 

\begin{table}[]
\begin{tabular}{|ccc|}
\hline
\multicolumn{3}{|c|}{\textbf{DeepTICA 1}}                                                                                                                                                                                 \\ \hline
\multicolumn{1}{|c|}{States}                      & \multicolumn{1}{c|}{\begin{tabular}[c]{@{}c@{}}Distances (99\%)\end{tabular}}  & \begin{tabular}[c]{@{}c@{}}Dihedrals (98\%)\end{tabular}  \\ \hline
\multicolumn{1}{|c|}{1 vs 2}                      & \multicolumn{1}{c|}{TYR1-O -- TYR10-N}                                                       & ASP3-$\chi_1$                                      \\ \hline
\multicolumn{3}{|c|}{\textbf{DeepTICA 2}}                                                                                                                                                                                 \\ \hline
\multicolumn{1}{|c|}{States}                      & \multicolumn{1}{c|}{\begin{tabular}[c]{@{}c@{}}Distances (97\%)\end{tabular}}   & \begin{tabular}[c]{@{}c@{}}Dihedrals (100\%)\end{tabular}  \\ \hline
\multicolumn{1}{|c|}{\multirow{2}{*}{2.1 vs 2.2}} & \multicolumn{1}{c|}{\multirow{2}{*}{THR6-Os -- THR8-Os}}                                     & THR6-$\chi_1$                                                          \\ \cline{3-3} 
\multicolumn{1}{|c|}{}                            & \multicolumn{1}{c|}{}                                                                        & THR8-$\chi_1$                                                          \\ \hline
\multicolumn{1}{|c|}{States}                      & \multicolumn{1}{c|}{\begin{tabular}[c]{@{}c@{}}Distances (100\%)\end{tabular}} & \begin{tabular}[c]{@{}c@{}}Dihedrals (100\%)\end{tabular} \\ \hline
\multicolumn{1}{|l|}{2.1 vs 2.3}                  & \multicolumn{1}{c|}{THR8-Os -- TRP9-N}                                                       & THR6-$\chi_1$                                                          \\ \hline
\end{tabular}
    \caption{
    Chignolin, unbiased simulation. Descriptors selected by the LASSO estimator for the chignolin system. Each row contains the result of the classifiers for the states identified by a given DeepTICA CV, either using the H-bond contact functions or dihedral angles. The percentages in parenthesis indicate the classification accuracy.
    }
    \label{tab:supporting_chignolin_unbiased_descriptors}
\end{table}

\section{BPTI, analysis with SASA descriptors}\label{app:bpti}

SASA contributions are calculated using the Shrake and Rupley algorithm~\cite{Shrake1973EnvironmentInsulin} as implemented in \verb_mdtraj_.

\begin{figure}
    \centering
    \includegraphics[width=\linewidth]{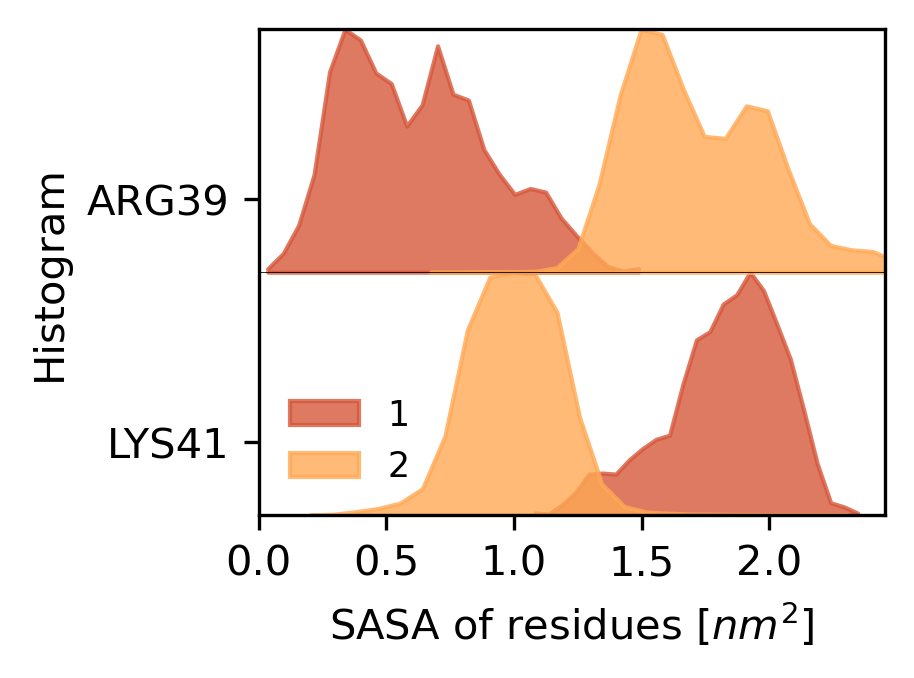}
    \caption{Histograms of the SASA features selected by the LASSO classifier in contrasting states 1 and 2. 
    }
    \label{fig:supporting_bpti_sasa}
\end{figure}

\begin{figure}[h!]
\centering

 \includegraphics[clip,width=1.\linewidth]{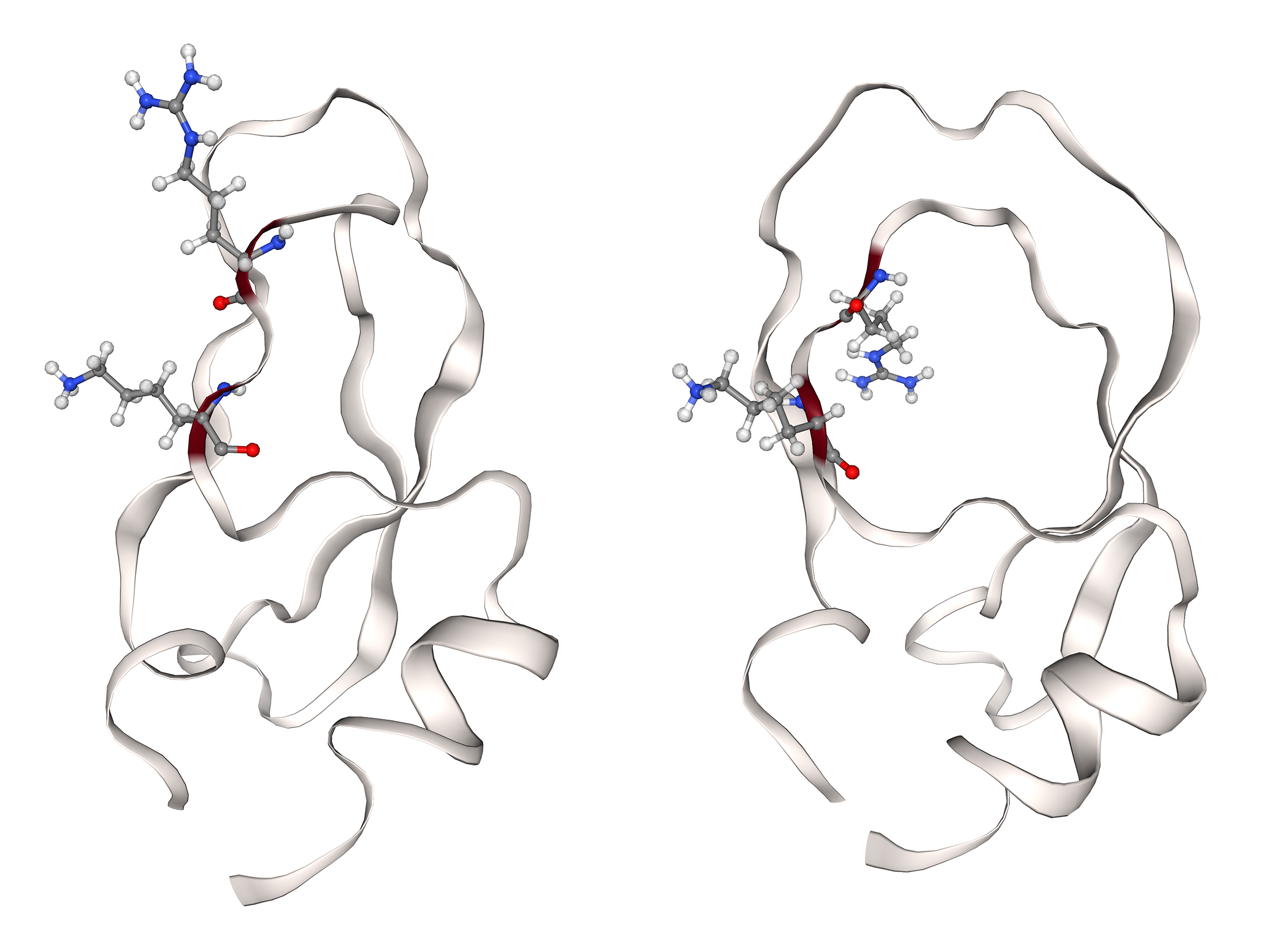}%

\caption{BPTI. Snapshots of state 1 (left) and crystal structure (right) to highlight the opening of the upper turns of the protein associated with state 1. The residues selected by the LASSO classifier with SASA contributions are highlighted in red and their atoms are displayed on top of the cartoon representation. }
\label{fig:supporting_bpti_snapshots}
\end{figure}

In Fig.~\ref{fig:supporting_bpti_sasa} we show the histograms of the selected residues, while in Fig.~\ref{fig:supporting_bpti_snapshots} we report a snapshot of the BPTI protein in state 1, compared with the crystallographic structure.

\end{document}